\documentclass[preprint,12pt]{elsarticle}

\usepackage[utf8]{inputenc}
\usepackage[T1]{fontenc}
\usepackage{graphicx}
\usepackage{amsmath,amssymb}
\usepackage{xcolor}
\usepackage[final]{changes}

\journal{Communications in Nonlinear Science and Numerical Simulation}

\providecommand{\cite}[1]{\citealp{#1}}

\begin{document}

\begin{frontmatter}

\title{Generalized flux-weighted boundary walls in kinetic models}

\author[mymainaddress]{Luca Barbieri}
\corref{mycorrespondingauthor}
\ead{luca.barbieri@obspm.fr}
\address[mymainaddress]{LIRA, Observatoire de Paris, Universit\'e PSL, Sorbonne Universit\'e, Universit\'e Paris Cit\'e, CY Cergy Paris Universit\'e, CNRS, 92190 Meudon, France}
\author[cnr,inaf,infn]{Pierfrancesco Di Cintio}
\ead{pierfrancesco.dicintio@cnr.it}
\address[cnr]{Istituto dei Sistemi Complessi, Consiglio Nazionale delle Ricerche (ISC-CNR), via Madonna del piano 10 50019, Sesto Fiorentino, Italy}
\address[inaf]{INAF -- Osservatorio Astrofisico di Arcetri Largo Enrico Fermi 5, 50125 Firenze, Italy}
\address[infn]{INFN -- Sezione di Firenze via Giovanni Sansone 1, 50019 Sesto Fiorentino, Italy.}

\begin{abstract}
We present a technique to investigate the stationary states of a system of a collisionless system confined by an external potential and coupled to boundary reservoirs through prescribed reinjection rules. We consider a family of boundary conditions parametrized by an integer $n$, corresponding to different velocity distributions imposed at the boundaries, generalizing the standard flux-weighted Maxwellian scheme.

By combining Liouville's theorem with the boundary injection rule, we derive an explicit analytical expression for the stationary distribution function. This framework provides a direct link between microscopic boundary dynamics and macroscopic stationary profiles. We show that thermal equilibrium is recovered only for the standard flux-weighted injection method, while for all other cases the system relaxes to manifestly non-thermal stationary states.

The resulting density and temperature profiles exhibit non-trivial spatial structures, including non-monotonic behaviour and temperature gradients induced by the boundary conditions alone. Analytical predictions for stationary moments are obtained in closed form for representative cases and are nicely reproduced by particle-based numerical simulations.
\end{abstract}


\end{frontmatter}

\section{Introduction}

Stationary configurations out of thermal equilibrium naturally occur in a wide variety of physical systems. Such configurations arise, for instance, in isolated particle systems with long-range interactions (i.e., mediated by two-body potentials decaying slower than $1/r^{-d}$ at large distances $r$, where $d$ is the spatial dimension \cite{Campa2014}). Relevant examples are self-gravitating systems \cite{binney2011galactic}, charged particle beams \cite{Levin2008,Teles-Levin2010}, vortex-vortex interactions in two-dimensional fluids \cite{BOUCHET2012227} and effective interactions mediated by electromagnetic fields \cite{Schuetz2014,Gupta2016}. In these systems, collisionless mean-field dynamics dominates over two-body encounters \cite{binney2011galactic}. As a consequence, the system develops increasingly fine structures in phase space, making macroscopic observables insensitive to small-scale variations and leading to a rapid evolution toward a stationary state, a process commonly referred to as violent relaxation \cite{Lynden-Bell1967,Kandrup_1998,Arad2005,Chavanis2006,Chavanis2006b,Beraldo2019,Barbieri2022,Teles2025,Beraldo2025}.

Out-of-thermal-equilibrium configurations also arise in non-isolated systems. A classical example is provided by transport phenomena \cite{Lepri2003,degl2019elementi,livi_politi_2017}, where gradients of physical quantities, such as temperature imposed by external reservoirs, drive the system toward non-equilibrium stationary states characterized by finite currents (e.g., heat fluxes in the presence of temperature gradients).

Plasma systems in contact with thermal or non-thermal boundaries constitute a particularly important subclass of such systems. These configurations are central to the study of solar and stellar atmospheres \cite{Scudder1992a,Scudder1992b,Barbieri2023temperature,barbieri2024temperaturedensityprofilescorona,Hau_2025,Barbieri2025c,Banik2026}, as well as to the acceleration and distribution functions of the solar wind \cite{Chamberlain1960,Brandt1966,Jockers1970,Lemaire1971,Maksimovic1997,Zouganelis2004,Vocks_2005,Pierrard2011,PtersdeBonhome2025,Vinogradov2026}. Similar boundary-driven kinetic effects also play a role in the modelling of fusion plasmas, in particular at the interface between the hot core and the scrape-off layer in TOKAMAK machines (e.g. see \cite{2005PPCF...47R.163F,2008CoPP...48...89T,Bufferand2013,Ciraolo2018,2026NucFu..66a6024M} and references therein).

Moreover, transport phenomena in particle systems in contact with thermal and non-thermal boundaries/baths have been widely studied in the context of out-of-equilibrium statistical mechanics, in particular in chains of coupled linear and non-linear oscillators (e.g. see \cite{Lepri1997,2000PhRvL..84.2144G,2007PhRvE..76e1118L,Lepri_2009,Lepri_2010,Iubini_2013,Iubini2017,Iubini2018,Andreucci2022,Andreucci2023}, for an extensive review see also \cite{2020FrP.....8..292B}).

Notably, \cite{ThermalWalls} demonstrated through $N$-particle simulations, first for an ideal gas and then for a system of hard spheres undergoing elastic collisions, that sampling particle velocities at the boundary from the mass-flux distribution drives the system toward thermal equilibrium, whereas sampling from a half-Maxwellian distribution does not.

Prompted by these findings, mass-flux sampling has been adopted in $N$-particle simulations with thermal or non-thermal boundaries in the context of plasma physics for different plasma parameters \cite{Landi-Pantellini2001,Landi2003,Landi_2012,Bercic2021,LepriMPC2021,Barbieri2024b,Barbieri2025b,Luo2025,Barbieri_2025c,Luo2025b}.

However, a general theoretical framework capable of predicting the stationary distribution function for different boundary injection prescriptions, and of demonstrating that only mass-flux sampling leads to a Maxwell-Boltzmann distribution while alternative prescriptions (such as sampling from a half-Maxwellian) produce non-thermal stationary states, is still lacking.

In this work, we address this issue by considering a simple kinetic model of collisionless particle systems confined in a one-dimensional potential and coupled to boundary reservoirs through a class of generalized reinjection rules. These rules are parametrized by an integer $n$ and include, as special cases, the standard flux-weighted Maxwellian injection ($n=1$) as well as other physically motivated prescriptions.

We develop a theoretical framework to determine the stationary distribution function of the system. By combining Jeans' theorem with the mass flux associated with the imposed boundary injection rule, we derive an explicit analytical expression for the stationary state, explicitly linking its form to the chosen boundary prescription. This approach establishes a direct connection between the microscopic boundary dynamics and the resulting macroscopic profiles.

Our analysis shows that thermal equilibrium is recovered only for the standard flux-weighted injection ($n=1$), while for all other values of $n$ the system reaches non-thermal stationary states. These states exhibit characteristic features, such as non-monotonic density profiles and spatially varying temperature, which are entirely induced by the boundary conditions.

The theoretical predictions are validated against direct $N$-particle simulations, showing excellent agreement across all cases considered. These results highlight the crucial role of boundary modelling in collisionless systems and provide a framework to analyze boundary-driven non-equilibrium stationary states.

The paper is organized as follows. In Section~\ref{sec:model}, we introduce the kinetic $N$-particle model and the class of boundary conditions considered. In Section~\ref{sec:theory}, we derive the analytical expression for the stationary distribution function. In Section~\ref{sec:numerics}, we present the numerical setup and compare the simulations with the theoretical predictions. Finally, in Section~\ref{sec:conclusions}, we summarize the main results of this work, highlighting both their novelties and limitations, and discuss the physical application that motivates the present study and comment on possible future perspectives of the proposed framework.

\section{The kinetic $N$-particle models}\label{sec:model}

We consider a one dimensional system of $N$ non-interacting particles confined in a finite domain $[a,b]$. The dynamics of each particle is governed by usual Hamilton equations
\begin{equation}\label{equationsofmotion}
\begin{cases}
\dot{q}_i = p_i \\
\dot{p}_i = -\dfrac{\partial U}{\partial q_i}
\end{cases}
\qquad \forall i=1,\dots,N \,.
\end{equation}
where $q_i$ and $p_i$ denote the position and momentum of the $i$-th particle within the domain, respectively, and $U(q)$ is an external potential such that $U(a)=U(b)=0$.

When a particle reaches the boundary (either $a$ or $b$) with momentum $p^\prime$, it is reinjected into the system with a new momentum $p$ such that ${\rm sign}(p)=-{\rm sign}(p^\prime)$, sampled from a family of distributions defined via the cumulative relation
\begin{equation}\label{montecarlo}
    r = A_{T,n} \int_0^p p^n e^{-\frac{p^2}{2T}} \, dp \,,
\end{equation}
where $r \in (0,1)$ is a uniformly distributed random variable and $T$ is the Maxwellian temperature. The normalization constant is
\begin{equation}
    A_{T,n} = 2^{\frac{1-n}{2}}
    T^{-\frac{n+1}{2}}
    \left(\Gamma\left(\frac{n+1}{2}\right)\right)^{-1},
\end{equation}
and $n$ is an integer parameter defining the family of boundary conditions.

For $n=0$ we recover the half-Gaussian distribution
\begin{equation}
    f(p) = \sqrt{\frac{2}{\pi T}} e^{-\frac{p^2}{2T}}, \qquad p \ge 0,
\end{equation}
which we refer to as a \emph{fixed thermal distribution}.

Rather than explicitly inverting Eq.~\eqref{montecarlo} in this case, the distribution is sampled using the Box--Muller transformation (e.g. \cite{cowan1998statistical}). Let $u_1,u_2 \in (0,1)$ be two independent uniformly distributed random variables. The Box-Muller transform generates a Gaussian random variable with zero mean and variance $T$:
\begin{equation}
    \tilde p = \sqrt{-2T\ln u_1}\,\cos(2\pi u_2).
\end{equation}
Since only momenta directed toward the interior are allowed, the reinjected momentum is taken as
\begin{equation}\label{boxmullerhalf}
    p = |\tilde p|
    =
    \left|\sqrt{-2T\ln u_1}\,\cos(2\pi u_2)\right|,
\end{equation}
which yields the required half-Gaussian distribution.

For $n \ge 1$, Eq.~\eqref{montecarlo} defines flux-weighted distributions. In particular, we focus on the physically relevant cases corresponding to the mass flux ($n=1$) and the energy flux ($n=3$), for which Eq.~\eqref{montecarlo} can be inverted analytically.

For $n=1$, one obtains
\begin{equation}\label{boundaryn1}
    p=\sqrt{-2T \ln(1-r)} \quad,
\end{equation}
while for $n=3$ the inversion yields
\begin{equation}\label{boundaryn3}
p =
\sqrt{
2T\left[
-1 - W_{-1}\!\left(-\frac{1-r}{e}\right)
\right]
} \quad.
\end{equation}

Here, $W_{-1}$ denotes the lower real branch of the Lambert $W$ function, defined as the solution $w \le -1$ of the
\begin{equation}
w e^{w} = x \quad,\qquad x \in [-1/e,0) \quad.
\end{equation}

Since there is no closed-form expression in terms of elementary functions, $W_{-1}(x)$ is evaluated numerically as the root of
\begin{equation}\label{fw}
f(w)=w e^{w}-x.
\end{equation}

The root is computed using Halley's method \cite{quarteroni2007numerical}
\begin{equation}
w_{n+1}
=
w_n
-
\frac{2 f(w_n) \frac{df}{dw_n}}
{2\left(\frac{df}{dw_n}\right)^2 - f(w_n) \frac{d^2f}{dw_n^2}},
\end{equation}
where $w_n$ denotes the $n$-th iterate of Halley's method.

The initial guess is constructed using asymptotic approximations of the Lambert $W_{-1}$ function in the relevant regions of its domain.

For $x \to 0^{-}$, we employ the leading-order asymptotic expansion
\begin{equation}
w_0 = \ln(-x) - \ln\!\bigl(-\ln(-x)\bigr),
\end{equation}
which provides an accurate approximation in this limit.

In the vicinity of the branch point $x=-1/e$, a different approximation is required due to the square-root singularity of the Lambert function $W_{-1}$. In this regime, we use a local asymptotic expansion in terms of the variable $y = x + 1/e$, namely
\begin{equation}
w_0 \approx -1 - \sqrt{2 e\, y} - \frac{1}{3}(2 e\, y),
\end{equation}
which captures the behavior of the function near the branch point.

In practice, a threshold value $x_{\mathrm{th}}$ is introduced to switch between these two initializations. This value is chosen empirically to ensure continuity of the initial guess and fast convergence of the iterative scheme over the entire interval $x \in [-1/e,0)$.

Starting from this initial guess, the solution is refined by iterating Halley's method. To ensure that the iteration remains on the $W_{-1}$ branch, any update that produces a value $w_{n+1} > -1$ is corrected by mapping it back into the region $w<-1$. The iteration is stopped when the absolute difference between two successive iterates satisfies
\begin{equation}
|w_{n+1} - w_n| < \varepsilon,
\end{equation}
where $\varepsilon$ is a fixed confidence threshold. Alternatively, a maximum number of $N_{\max}$ iterations can be also imposed. In practice, the parameters $\varepsilon$ and $N_{\max}$ control the accuracy and computational cost of the procedure, respectively.

The robustness of the numerical solution with respect to the choice of the threshold $x_{\mathrm{th}}$ and the convergence parameters ($\varepsilon$, $N_{\max}$) has been explicitly verified. No significant variations in the computed values were observed over a wide range of these parameters.

\section{Evaluation of the stationary state distribution function}\label{sec:theory}

Starting from a generic initial condition, the system is driven toward a stationary state by the interaction with the boundary. Since the particles are non-interacting, the corresponding phase-space distribution function $f(q,p)$ obeys the stationary Vlasov equation
\begin{equation}\label{StationaryVlasov}
    \{H,f\} = 0,\qquad H(q,p)=\frac{p^2}{2}+U(q),
\end{equation}
where $H$ is the single-particle Hamiltonian and $
\{\cdot,\cdot\}$ denotes the standard Poisson bracket. This condition implies that $f$ is constant along the characteristics of the Hamiltonian flow (e.g. \cite{goldsteinclassical}). In one-dimensional integrable systems, the single-particle energy $H$ is the only invariant of motion, and therefore, in virtue of Jeans' theorem \cite{nicholson1983introduction}, the stationary distribution can be written as
\begin{equation}\label{jeanstheorem}
    f(q,p)=F(H).
\end{equation}

The function $F(H)$ is determined by the boundary injection rule. It is convenient to express such a rule in terms of the incoming particle mass-flux. By construction, the cumulative mass-flux of the particles reinjected with momentum between $0$ and $p$ is
\begin{equation}
    J_{p>0}=\int_0^p p'\,\tilde f_n(p')\,dp',
\end{equation}
where
\begin{equation}
    \tilde f_n(p)=A_{T,n}p^{n-1}e^{-p^2/(2T)} \quad,\qquad p>0.
\end{equation}
Accordingly, the differential incoming flux is
\begin{equation}
    dJ = p\,\tilde f_n(p)\,dp
    = A_{T,n}p^{n}e^{-p^2/(2T)}\,dp,
\end{equation}
which reproduces the boundary distribution introduced in Eq.~\eqref{montecarlo}.

At the kinetic level, the flux of particles crossing the boundary with momentum in $(p,p+dp)$ is given by $p\,f(q=a;b,p)\,dp$. Imposing that the incoming flux at the boundary coincides with the prescribed injection rule, one obtains
\begin{equation}
    p\,f(q=a;b,p)\,dp = p\,\tilde f_n(p)\,dp, \qquad p>0,
\end{equation}
from which it follows that
\begin{equation}
    f(q=a;b,p)=\tilde f_n(p),\qquad p>0.
\end{equation}

This condition specifies the distribution function only for incoming particles at the boundary. However, since the Hamiltonian is even in $p$, i.e. $H(q,p)=H(q,-p)$, the stationary distribution $f$ inherits the same symmetry because of Eq. \eqref{jeanstheorem} and therefore satisfies
\begin{equation}
    f(q,p)=f(q,-p).
\end{equation}
As a consequence, once the distribution is fixed for $p>0$ at the boundary, it is also uniquely determined for $p<0$, allowing one to extend the boundary condition to the entire phase space.

For a phase-space point $(q,p)$ with energy $H$, the corresponding characteristic intersects the boundary with an incoming momentum $p_{a;b}>0$ such that
\begin{equation}
    p_{a;b}=\sqrt{2H}.
\end{equation}
Since $f$ is constant along the characteristics, the value of the distribution at $(q,p)$ is fixed by its value at the boundary along the same trajectory. Therefore,
\begin{equation}
    F(H)=\tilde f_n(p_{a;b})
    =A_{T,n}\,(2H)^{\frac{n-1}{2}}e^{-\frac{H}{T}}.
\end{equation}
Up to a multiplicative constant independent of $H$, the stationary distribution thus reads
\begin{equation}
    f_n(q,p)\propto H^{\frac{n-1}{2}}e^{-H/T}.
\end{equation}

Imposing the normalization condition
\begin{equation}
    \int_{a}^{b}dq\int_{-\infty}^{+\infty}dp\;
     f_n(q,p) = 1 \quad,
\end{equation}
the stationary distribution reads
\begin{equation}\label{VDF}
    f_n(q,p)=
    \frac{
    H^{\frac{n-1}{2}}e^{-H/T}
    }{
    \displaystyle
    \int_{a}^{b}dq\int_{-\infty}^{+\infty}dp\;
    H^{\frac{n-1}{2}}e^{-H/T}
    } \quad.
\end{equation}

As a consequence, the stationary distribution reduces to the equilibrium Maxwell-Boltzmann form only for $n=1$, corresponding to the standard flux-weighted Maxwellian injection. For $n\neq 1$, the system reaches genuinely non-thermal stationary states.

This result provides a theoretical explanation of the numerical findings reported in \cite{ThermalWalls}, where it was shown that an ideal gas in contact with a thermal boundary reaches thermal equilibrium only when the injection mechanism corresponds to the flux-weighted distribution with $n=1$. In that work, the cases $n=0$ and $n=1$ were investigated numerically. Here, we extend those results by deriving an explicit analytical expression for the stationary distribution, valid for arbitrary values of $n$ and for a general external potential $U(q)$.

\section{Numerical validation of the theory}\label{sec:numerics}

\subsection{Numerical setup and evaluation of macroscopic observables}
Let us consider the simple system of non-interacting particles evolving in the external potential
\begin{equation}\label{model1}
    U(q) = 2 \cos{\left(\frac{q}{2}\right)} \quad, \qquad q \in [-\pi,\pi] \quad.
\end{equation}

The latter is a dimensionless formulation of the single-species external potential employed in \cite{Barbieri2023temperature,Barbieri2024b} to model collisionless coronal-loop plasmas in stellar atmospheres. More precisely, it represents the dimensionless counterpart of the total potential used in those works, obtained as the sum of the solar gravitational potential and the large-scale electrostatic potential generated by the stellar interior. The latter corresponds to the Pannekoek--Rosseland electric field \citep{Pannekoek_1922,Rosseland_1924,Neslusan2001-rp,belmont2013collisionless}, which counterbalances the gravitational separation of charges and thereby ensures the quasi-neutrality of the atmospheric plasma. As a result, both particle species experience the same total potential, whose dimensionless form is given by Eq.~\eqref{model1} \citep{Barbieri2023temperature,Barbieri2024b}.

In that context, the potential is expressed as a function of the curvilinear coordinate along a semicircular magnetic flux tube representing a coronal loop. The trigonometric dependence appearing in Eq.~\eqref{model1} originates from this geometrical parametrization. Accordingly, the variable $q$ used in the present work should be interpreted as the dimensionless curvilinear coordinate measured along the loop.

Within this physical framework, the single-particle Hamiltonian considered here can be interpreted as the Hamiltonian of an electron moving along the curvilinear coordinate $q$ of the loop with conjugate momentum $p$. The corresponding proton Hamiltonian has the same effective potential but a kinetic-energy term rescaled by the proton-to-electron mass ratio, as discussed in \cite{Barbieri2023temperature,Barbieri2024b}.

The collisionless dynamics is obtained by integrating the equations of motion given by Eq.~\eqref{equationsofmotion} with the potential \eqref{model1}. Due to the Hamiltonian structure of the system, we employ an explicit fourth-order symplectic integrator \cite{Candy1991}, ensuring a global error scaling as $dt^4$, where $dt$ is the (fixed) integration time step.

The boundary conditions are implemented using Eq.~\eqref{boxmullerhalf} for $n=0$, Eq.~\eqref{boundaryn1} for $n=1$, and Eq.~\eqref{boundaryn3} for $n=3$.

In all all numerical experiments we set the initial conditions in thermal equilibrium. Positions and momenta are sampled using a von Neumann acceptance-rejection algorithm \cite{press2007} from the equilibrium distribution
\begin{equation}
    f_{eq}(q,p)= 
    \frac{e^{-H}}{
    \displaystyle
    \int_{-\infty}^{+\infty} dp 
    \int_{-\pi}^{\pi} dq \, e^{-H}
    } \quad.
\end{equation}

To compare theoretical predictions with numerical simulations, we compute the first two velocity moments of the distribution function defined in Eq.~\eqref{VDF} which is satisfied by the potential given by Eq. ~\eqref{model1}. The number density, as usual, is given by
\begin{equation}
    n_n(q) = \int_{-\infty}^{+\infty} f_n(q,p)\, dp,
\end{equation}
while the kinetic temperature reads
\begin{equation}
    T_n(q) = \frac{\int_{-\infty}^{+\infty} p^2 f_n(q,p)\, dp}{\int_{-\infty}^{+\infty} f_n(q,p)\, dp}.
\end{equation}

For $n=0$, momentum integrals can be expressed in terms of modified Bessel functions of the second kind. We obtain
\begin{equation}\label{densityn0}
    n_0(q)=
    \frac{
    e^{-U(q)/(2T)}K_0\!\left(\frac{U(q)}{2T}\right)
    }
    {
    \displaystyle \int_{-\pi}^{\pi}
    e^{-U(q')/(2T)}K_0\!\left(\frac{U(q')}{2T}\right)\,dq'
    },
\end{equation}
and
\begin{equation}\label{temperaturen0}
    T_0(q)=
    U(q)\left[
    \frac{K_1\!\left(\frac{U(q)}{2T}\right)}{K_0\!\left(\frac{U(q)}{2T}\right)}-1
    \right].
\end{equation}

Here $K_\nu(x)$ denotes the modified Bessel function of the second kind of order $\nu$, defined in integral form as
\begin{equation}
    K_\nu(x) = \int_0^{+\infty} e^{-x \cosh t} \cosh(\nu t)\, dt \quad, \qquad x>0 \,.
\end{equation}
In particular,
\begin{equation}
    K_0(x) = \int_0^{+\infty} e^{-x \cosh t} \, dt \quad, 
    \qquad
    K_1(x) = \int_0^{+\infty} e^{-x \cosh t} \cosh t \, dt \quad.
\end{equation}
For for our choice of external potential, one has
\begin{equation}
U(q)\sim q+\pi \qquad \text{for } q\to -\pi^+.
\end{equation}
Using the small-argument expansions of the modified Bessel functions,
\begin{equation}
K_0(x)\sim -\ln\!\left(\frac x2\right)-\gamma,
\qquad
K_1(x)\sim \frac1x,
\qquad x\to 0^+,
\end{equation}
where $\gamma \sim 0.5772$ is the Euler-Mascheroni constant, one finds for the density profile
\begin{equation}
n_0(q)\sim
-\frac{\ln\!\left(\frac{U(q)}{4T}\right)+\gamma
}{
\displaystyle \int_{-\pi}^{\pi}
e^{-U(q')/(2T)}K_0\!\left(\frac{U(q')}{2T}\right)\,dq'
},
\qquad q\to -\pi^+.
\end{equation}
Therefore,
\begin{equation}\label{densityn0asymp}
n_0(q)\sim -\ln(q+\pi),
\qquad q\to -\pi^+,
\end{equation}
showing that the density diverges logarithmically at the boundary.

Similarly, for the temperature profile one obtains
\begin{equation}
T_0(q)\sim
-\frac{2T}{\ln\!\left(\frac{U(q)}{4T}\right)+\gamma},
\qquad q\to -\pi^+,
\end{equation}
and hence
\begin{equation}\label{temperaturen0asymp}
T_0(q)\sim
-\frac{2T}{\ln(q+\pi)},
\qquad q\to -\pi^+.
\end{equation}
Thus, while the density diverges logarithmically, the kinetic temperature vanishes logarithmically at the boundary.

In contrast, for $q\to 0$ the potential $U(q)$ admits the expansion
\begin{equation}
U(q)=2-\frac{q^2}{4}+O(q^4).
\end{equation}
Using this expression, the density profile can be expanded as
\begin{equation}\label{densityn0q0}
n_0(q)
=
\frac{e^{-1/T}K_0(1/T)}{\mathcal N_0}
+
\frac{e^{-1/T}}{8T\,\mathcal N_0}
\bigl[K_0(1/T)+K_1(1/T)\bigr]\,q^2
+O(q^4),
\end{equation}
where the normalization constant $\mathcal N_0$ is given by
\begin{equation}
\mathcal N_0=
\int_{-\pi}^{\pi}
e^{-U(q')/(2T)}K_0\!\left(\frac{U(q')}{2T}\right)\,dq'.
\end{equation}
Hence, in the vicinity of $q=0$, the density behaves as a positive quadratic function.

The corresponding temperature profile admits the expansion
\begin{equation}\label{temperaturen0q0}
T_0(q)
=
2\left[\frac{K_1(1/T)}{K_0(1/T)}-1\right]
-\frac{q^2}{4}
\left[
\frac{1}{T}\left(\frac{K_1(1/T)}{K_0(1/T)}\right)^2
-\frac{1}{T}
-1
\right]
+O(q^4),
\end{equation}
corresponding instead to negative quadratic function. The absence of the linear terms in both expansions reflects the quadratic behaviour of the potential around $q=0$, which implies a locally symmetric structure of the stationary state.

For the $n=1$ case, one obtains
\begin{equation}\label{densityandtemperature1}
    n_1(q)=
    \frac{e^{-U(q)/T}}
    {\displaystyle \int_{-\pi}^{\pi}e^{-U(q')/T}\,dq'},
    \qquad
    T_1(q)=T,
\end{equation}
While for $n=3$, one finds
\begin{equation}\label{densityn3}
    n_3(q)=
    \frac{\left(U(q)+\frac{T}{2}\right)e^{-U(q)/T}}
    {\displaystyle \int_{-\pi}^{\pi}\left(U(q')+\frac{T}{2}\right)e^{-U(q')/T}\,dq'},
\end{equation}
and
\begin{equation}\label{temperaturen3}
    T_3(q)=
    T\,\frac{U(q)+\frac{3}{2}T}{U(q)+\frac{1}{2}T}.
\end{equation}
To monitor the dynamical evolution of the system, we compute the total average kinetic energy
\begin{equation}\label{kineticenergy}
    K=\frac{1}{N}\sum_{j=1}^{N}\frac{p_j^2}{2}.
\end{equation}
The system is considered to have reached the stationary state once $K$ fluctuates around a constant value within a given confidence width. In the stationary regime, the number density and the kinetic temperature are computed at different time snapshots and subsequently averaged in time. The resulting numerical profiles are then compared with the corresponding analytical predictions for the three boundary conditions.

\subsection{Comparison of numerical results with theory}

\begin{figure}
    \centering
    \includegraphics[width=0.99\columnwidth]{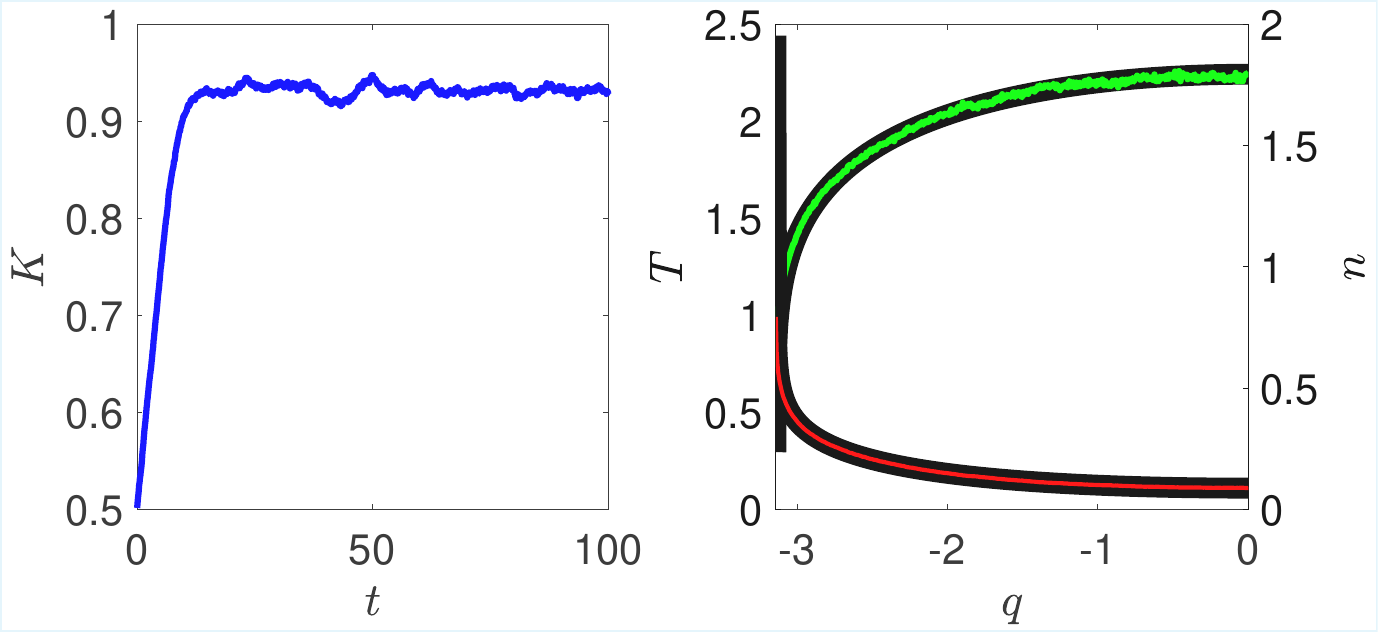}
    \caption{
    Left panel: time evolution of the total kinetic energy, computed from Eq.~\eqref{kineticenergy}, showing the approach to the stationary state. 
    Right panel: stationary density profile (red) and kinetic temperature profile (green) as functions of $q \in [-\pi,0]$ (only half of the interval is shown, since the potential is symmetric with respect to $q=0$), obtained from numerical simulations of Eq.~\eqref{equationsofmotion} with the potential $U(q)$ given by Eq.~\eqref{model1}. The boundary condition corresponds to the case $n=0$, implemented via Eq.~\eqref{boxmullerhalf}. 
    The corresponding analytical predictions, computed from Eq.~\eqref{densityn0} for the density and Eq.~\eqref{temperaturen0} for the temperature are marked by the heavy black lines. 
    Simulation parameters: $T=3$, $N=2^{16}$, and $dt=10^{-2}$, $x_{\mathrm{th}} \simeq -0.30$, $\varepsilon = 10^{-12}$,$N_{\max}=50$.
    }
    \label{fig1}
\end{figure}

\begin{figure}
    \centering
    \includegraphics[width=0.99\columnwidth]{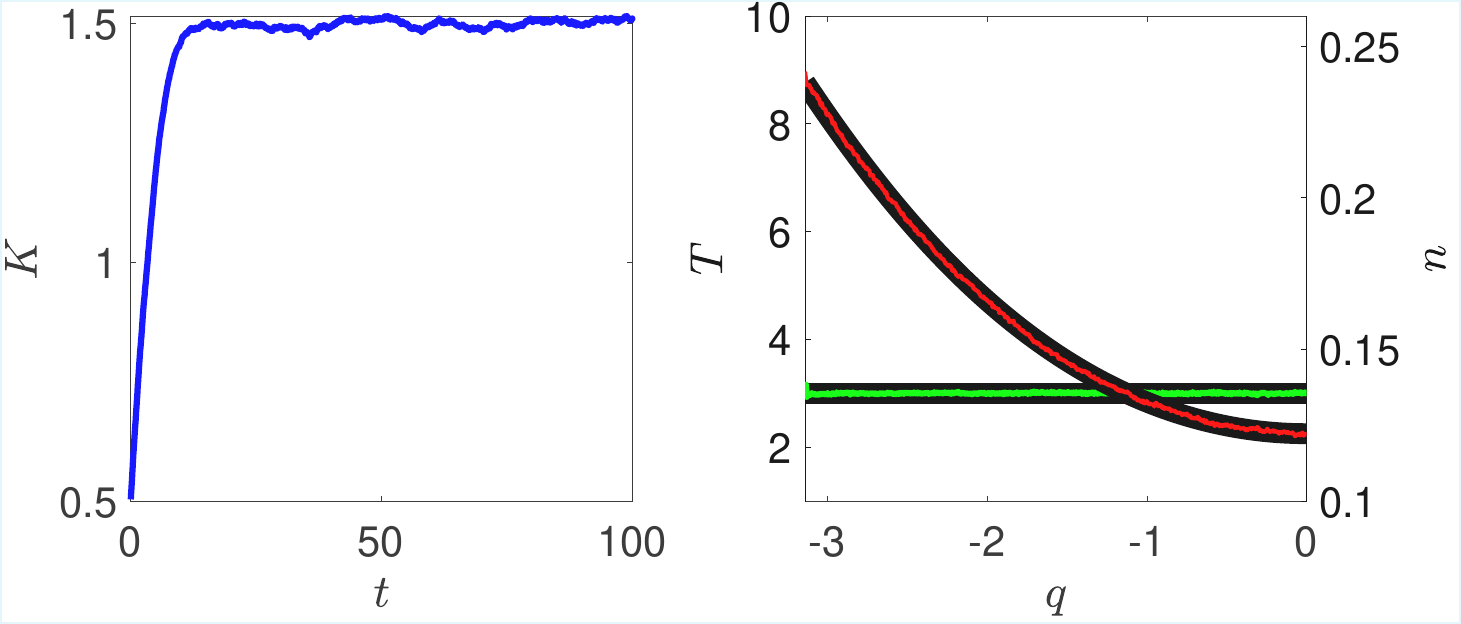}
    \caption{
    Same quantities as in Fig.~\ref{fig1}, with the same colour coding and parameters. In this case, the boundary condition corresponds to mass flux ($n=1$), implemented via Eq.~\eqref{boundaryn1}. The analytical density and temperature profiles are computed from Eqs.~\eqref{densityandtemperature1}.
    }
    \label{fig2}
\end{figure}

\begin{figure}
    \centering
    \includegraphics[width=0.99\columnwidth]{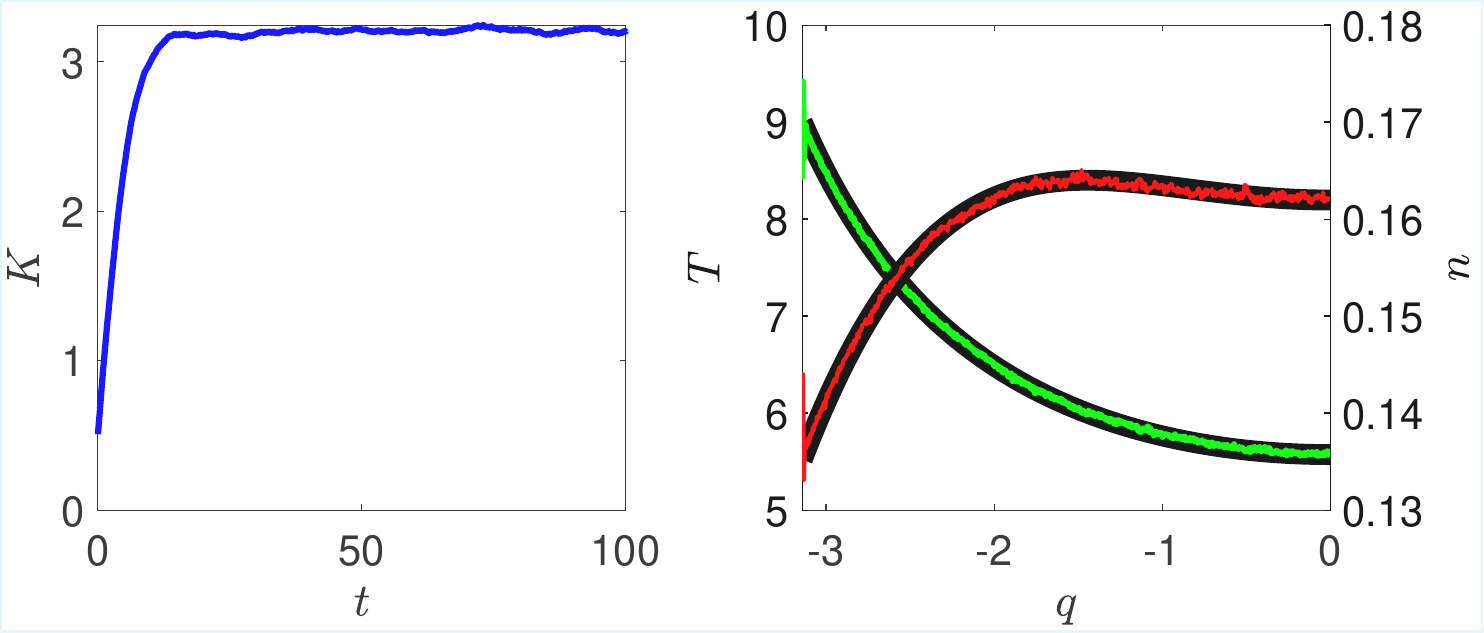}
    \caption{
    Same quantities as in the previous figures, with the same colour coding and parameters. In this case, the boundary condition corresponds to energy flux ($n=3$), implemented via Eq.~\eqref{boundaryn3}. The analytical density and temperature profiles are computed from Eqs.~\eqref{densityn3} and \eqref{temperaturen3}, respectively.
    }
    \label{fig3}
\end{figure}

The results for the different choices of the boundary conditions are shown in Figs.~\ref{fig1}, \ref{fig2} and \ref{fig3}, respectively. As shown in the left panels of all figures, the system always relaxes on a stationary state. Note that in all the right panels, due to the symmetry with respect to $q=0$ we show only half of the spatial domain. Overall, we observe in all cases a remarkably good agreement of the numerical simulations with the theoretical predictions.

We first consider the $n=0$ case (Fig.~\ref{fig1}), corresponding to the boundary condition defined by Eq.~\eqref{boxmullerhalf}. The system's non-thermal stationary-state is characterized by a decreasing density profile and an increasing temperature profile as functions of $q \in [-\pi,0]$. 

The sharp accumulation of particles near the left boundary is explained by the asymptotic behaviour of the density and temperature profiles, as given by Eqs.~\eqref{densityn0asymp} and \eqref{temperaturen0asymp}. In particular, the logarithmic divergence of the density described by Eq.~\eqref{densityn0asymp} is accompanied by a logarithmically vanishing kinetic temperature \eqref{temperaturen0asymp}, indicating that the boundary singularity corresponds to an increasingly populated but colder region. This strong anti-correlation between the density (decreasing) and temperature (increasing) profiles is progressively smoothed as one approaches $q=0$. This is consistent with the Taylor expansions of the density and temperature profiles around $q=0$, given by Eqs.~\eqref{densityn0q0} and \eqref{temperaturen0q0}, which show that both profiles are locally quadratic functions of $q$ with opposite curvature, exhibiting a minimum (for the density) and a maximum (for the temperature) at $q=0$.

In the $n=1$ case (Fig.~\ref{fig2}), corresponding to the boundary condition of mass flux given by Eq.~\eqref{boundaryn1}, the system relaxes toward a thermal equilibrium state. Here, Eq.~\eqref{VDF} reduces to the Maxwell-Boltzmann distribution, implying an isothermal temperature profile and a density profile decreasing with $q$, consistent with the increasing trend of the potential $U(q)$.

Finally, for the $n=3$ case (Fig.~\ref{fig3}), corresponding to the boundary condition of the energy flux defined by Eq.~\eqref{boundaryn3}, the system again reaches a stationary state. However, for such choice the density profile is non-monotonic. This becomes evident from Eq.~\eqref{VDF}, where the distribution function is proportional to $H e^{-H/T}$. In practice, the stationary density (see Eq.~\eqref{densityn3}) is the product of an increasing contribution (due to the factor $H$) and a decreasing exponential factor. Depending on the parameters, the increasing contribution may dominate, leading to a larger density at higher values of $q$. The corresponding temperature profile is instead a decreasing function of $q$ in the interval $q \in [-\pi,0]$, in agreement with Eq.~\eqref{temperaturen3}, which explains the monotonic behaviour observed in Fig.~\ref{fig3}.

\section{Summary, discussions and perspectives}\label{sec:conclusions}

In this work, we have investigated the stationary states of a collisionless system of non-interacting particles confined in a one-dimensional potential and coupled to boundary reservoirs through prescribed reinjection rules. Considering a family of boundary conditions parametrized by an integer $n$, we have shown how the statistical properties of the stationary state are directly determined by the nature of the imposed boundary flux.

We have derived an explicit analytical expression (Equation \eqref{VDF}) for the stationary distribution function by combining Jeans' theorem with the boundary flux imposed by the injection rule. This approach provides a unified framework linking microscopic dynamics to macroscopic stationary profiles and remains valid for arbitrary values of $n$ and for a general external potential.

Our analysis shows that thermal equilibrium is recovered only in the case $n=1$, which corresponds to a Maxwellian injection of mass flux. For all other values of $n$, the system reaches non-thermal stationary states by construction. In particular, for $n=0$ we observe temperature profiles increasing with $q$, while for $n=3$ the competition between energetic and thermal contributions leads to non-monotonic density profiles and decreasing temperature profiles.

Our analytical predictions are nicely supported by numerical simulations of a simple one dimensional model, for which for all the explored boundary conditions (i.e. $n=0,1,3$) induce the expected temperature and density profiles.

We note that, the crucial role of the particle velocities sampling scheme from the mass flux distribution in driving the system toward thermal equilibrium, was already uncovered through $N$-particle simulations in \cite{ThermalWalls}. Our results provide a clear theoretical interpretation of these findings and extend them to a broader class of boundary conditions and external potentials. Moreover, we note also that, the technique discussed here remains valid even in the case where the confining nature of the potentials is relaxed (i.e., we allow particle escape) provided that the bulk dynamics remains collisionless, such as for example, stellar or planetary atmospheres in their supporting gravitational fields. 

All the results presented in this work have been obtained by considering Gaussian boundary distributions. However, the theoretical framework developed here is in fact more general and can be extended beyond this specific case. In particular, replacing the Gaussian weight $e^{-\frac{p^2}{2T}}$ in Eq.~\eqref{montecarlo} with a generic distribution $g(p)$ leads to a corresponding modification of the stationary distribution function, which can be expressed in terms of the single-particle energy as a function of $g(H)$ in the same fashion as Eq. \eqref{VDF}. 

We recall that, a collisionless kinetic $N$-particle two-component plasma model of the solar atmosphere was recently introduced in \cite{Barbieri2023temperature}. Moreover, a coarse-graining analytical framework was developed to study and interpret the stationary states emerging from the numerical simulations of that model \cite{Barbieri2024b,Barbieri2025b}.

In this series of papers, the plasma is driven out of equilibrium by a sequence of intermittent heating events occurring at different temperatures. Such heating pulses are modelled through the mass-flux injection technique originally introduced in \cite{ThermalWalls}. However, boundary heating can be implemented either through a mass flux or through an energy flux, the latter corresponding to the use of Eq.~\eqref{montecarlo} with $n=3$ instead of $n=1$. For this reason, in the present study we focus on the simplest configuration of a boundary characterized by a time-independent Gaussian distribution with fixed temperature. In the prosecution of the present work, we plan to apply the generalized family of boundary conditions introduced here to the two-component plasma atmosphere model \cite{Barbieri2023temperature} and to extend the temporal coarse-graining formalism \cite{Barbieri2024b,Barbieri2025b} accordingly.

We note that, a related boundary-driven collisionless plasma problem including self-consistent mean-field effects was investigated by \cite{Rizzato2009}, for the case of the stationary state of an electron beam confined between emitting and collecting electrodes, using the standard stationary Vlasov-Poisson framework. As a consequence, the electrostatic potential was not prescribed \emph{a priori}, but emerged from the space-charge distribution generated self-consistently.

The present approach differs in an important respect. Here, the confining potential is externally assigned and the stationary distribution is determined solely by the combination of collisionless Liouville transport and the prescribed boundary reinjection rule. This allows the stationary distribution function to be derived analytically in closed form for a broad class of boundary conditions. In contrast, the self-consistent problem considered in ~\cite{Rizzato2009} requires the simultaneous determination of both the distribution function and the mean electrostatic potential.

Despite these differences, the two approaches share a common physical foundation. In both cases the stationary state is governed by collisionless Vlasov dynamics and is strongly influenced by the particle distribution imposed at the boundaries. The results of \cite{Rizzato2009} show that self-consistent space-charge effects can generate strongly non-uniform density and temperature profiles even in the absence of binary collisions, consistently with the non-thermal stationary states obtained here for boundary conditions different from the standard mass-flux prescription.

The present formalism can in principle be generalized to include self-consistent mean-field effects. However, the main application motivating the present manuscript is the kinetic modelling of collisionless plasma atmospheres. In such systems, the stellar interior generates the well-known Pannekoek-Rosseland electric field \cite{Pannekoek_1922,Rosseland_1924,Neslusan2001-rp,belmont2013collisionless}, which counterbalances the gravitational separation of charges and thereby ensures the quasi-neutrality of the plasma in a gravitationally stratified atmosphere. For this reason, no significant self-consistent space-charge potential is expected to develop in the stationary state, and the dynamics can be accurately described in terms of particles evolving in an externally assigned confining potential.

Nevertheless, in more general situations involving self-consistent mean-field interactions, the effective potential is not prescribed externally but is generated by the particle distribution itself. In such cases, the stationary distribution function obtained through a Liouville mapping of the boundary distribution depends on the self-consistent potential, while the latter depends on the full distribution function through the corresponding mean-field equation. As a consequence, the determination of the stationary state becomes a nonlinear self-consistent problem.

Furthermore, the self-consistent potential may generate regions of phase space that are not dynamically connected to the boundaries. Particles located in such regions can remain trapped and therefore cannot be described directly by the boundary-driven stationary-state construction developed in this work. While the Liouville mapping remains applicable to the subset of particles whose trajectories remain connected to the boundaries, the trapped and boundary-connected populations both contribute to the self-consistent potential and are therefore intrinsically coupled. A complete determination of the stationary state consequently requires the simultaneous evaluation of both populations together with the associated mean field.

For electrostatic one-dimensional plasmas, this situation corresponds to the coupled Vlasov--Poisson problem where the electrostatic potential is generated self-consistently by the particle density. More generally, analogous difficulties arise in any collisionless system with long-range mean-field interactions. The trapped component is expected, under sufficiently efficient phase-space mixing, to relax toward a quasi-stationary or stationary state through the process of violent relaxation, and its asymptotic properties may therefore be investigated using the stationary-state theories developed in the works cited at the beginning of this Introduction.

Therefore, in the most general case, the procedure defined in the present work provides a complete determination of the stationary distribution function only when the effective single-particle potential, arising from the external potential, the self-consistent mean field, or their combination, generates phase-space trajectories that remain connected to the boundaries. Whenever disconnected trapped regions are present, additional information is required to determine the corresponding trapped component of the distribution function and to close the self-consistent problem.

\section*{Acknowledgments}
The authors wish to thank the anonymous reviewer for the valuable comments and suggestions that helped improve the clarity and quality of this work. L.B. wants to thank the Sorbonne Universit\'e in the framework of the Initiative Physique des Infinis for financial support. PF.D.C. acknowledges support from the MUR PRIN2022 project “Breakdown of
ergodicity in classical and quantum many-body systems”
(BECQuMB) Grant No. 20222BHC9Z.

\section*{Declaration of competing interest}
The authors declare that they have no known competing financial interests or personal relationships that could have appeared to influence the work reported in this paper.

\bibliographystyle{elsarticle-num}
\bibliography{jpp-instructions}

\end{document}